\documentclass[aps,prb,twocolumn,superscriptaddress,showpacs]{revtex4-2}
\usepackage[colorlinks,linkcolor=blue,citecolor=blue,urlcolor=blue,breaklinks=true]{hyperref}
\usepackage{amsmath,amssymb,graphics,epsfig,epstopdf,color,verbatim,braket,tabularx,breakurl}
\usepackage{multirow}
\usepackage{diagbox}
\usepackage{subfig}
\usepackage[normalem]{ulem}
\usepackage{float}
\usepackage{ragged2e}
\usepackage{placeins}

\begin{document}

\title{Probing universal imaginary-time relaxation critical dynamics with infinite projected entangled pair states}

\author{He-Yu Lin}
\address{School of Science, Inner Mongolia University of Science and Technology, Baotou 014010, China}
\address{Department of Physics, Renmin University of China, Beijing 100872, China}
\address{Key Laboratory of Quantum State Construction and Manipulation (Ministry of Education), Renmin University of China, Beijing, 100872, China}
\author{Shuai Yin}
\address{Guangdong Provincial Key Laboratory of Magnetoelectric Physics and Devices, Sun Yat-Sen University, Guangzhou, 510275, China}
\address{School of Physics, Sun Yat-Sen University, Guangzhou, 510275, China}
\author{Z. Y. Xie}
\email{qingtaoxie@ruc.edu.cn}
\address{Department of Physics, Renmin University of China, Beijing 100872, China}
\address{Key Laboratory of Quantum State Construction and Manipulation (Ministry of Education), Renmin University of China, Beijing, 100872, China}
\author{Zhong-Yi Lu}
\email{zlu@ruc.edu.cn}
\address{Department of Physics, Renmin University of China, Beijing 100872, China}
\address{Key Laboratory of Quantum State Construction and Manipulation (Ministry of Education), Renmin University of China, Beijing, 100872, China}
\date{\today}

\begin{abstract}
We investigate the imaginary-time relaxation critical dynamics of the two-dimensional transverse-field Ising model using infinite projected entangled pair states (iPEPS) with the full-update strategy. Simulating directly in the thermodynamic limit, we explore the relaxation process near the critical point with two types of initial states: a fully polarized state and a product state with a small magnetization. For the fully polarized state, the magnetization shows a power law scaling $M\propto \tau^{-\beta/(\nu z)}$ in the imaginary-time evolution, from which both the critical point and critical exponent can be determined with high accuracy. For the nearly paramagnetic state, the relaxation process exhibits a behavior of $M\propto \tau^\theta$ with $\theta=0.1958$ being the critical initial-slip exponent, which is in good agreement with that obtained from the dynamic scaling of the self-correlation in quantum Monte Carlo method. These universal features emerge well before the system converges to the ground state, demonstrating the efficiency of imaginary-time evolution for probing quantum criticality. Our results demonstrate that iPEPS can serve as a robust and scalable method for studying dynamical critical phenomena in two-dimensional quantum many-body systems.

\end{abstract}

\maketitle
\section{Introduction}
Understanding nonequilibrium critical dynamics is a central issue in statistical mechanics and condensed matter physics. Among various realizations of nonequilibrium processes, imaginary-time evolution has emerged as a robust framework that not only works as a routine method to determine the ground state of quantum many-body systems \cite{Vidal-TEBD, JHC2007, CorbozSC, PESS2014, Chang2024} but also encodes universal information in its transient stages \cite{Yin2014, Zhang2014, Shu2017, Li2023, Shu2022, Shu2022b}. Moreover, the study of imaginary-time dynamics is heating up due to recent advances in quantum computers \cite{Motta2020, Tsuchimochi2021}. 

Near quantum critical points, the imaginary-time evolution of a quantum system can exhibit universal scaling behaviors incorporating the equilibrium critical information of the quantum phase transitions and the intrinsic nonequilibrium scaling properties \cite{ Yin2014, Zhang2014, Shu2017, Li2023}, and provides an analogy to the short-time critical dynamics in classical critical points \cite{Janssen1989, Li1995, Zheng1996, Zheng1998}. This makes the imaginary-time evolution especially suitable for probing critical properties in strongly correlated systems \cite{Yin2014, Shu2022, Shu2022b}. A remarkable advantage of this method is that the computational cost can be significantly reduced, since universal critical properties can be reliably extracted during the short-time stage of the evolution process  \cite{Yin2014, Li1995}. Recently, this method has even shown its power in sign-problematic fermionic models in quantum Monte Carlo simulations \cite{Yu2024, ChangWX2023}. In addition, the scaling theory of the imaginary-time relaxation critical dynamics has been verified in programmable quantum circuits \cite{Zhang2024}. 

A series of numerical methods have been developed to study the critical properties of the strongly correlated many-body systems \cite{SCSBook2013}. Among these efforts, tensor network states combined with the renormalization group techniques \cite{TXBook2023} are free of sign problems, can deal with the infinite quantum systems satisfying the area law of the entanglement entropy \cite{RMP2010}, and thus are drawing increasing attentions in studying quantum magnetization \cite{CorbozSSM, Kagome2017, QianLi2022, NXSSM, LHY2024}, superconductivity \cite{CorbozSC, YS2023}, topological order \cite{MJW2017, CJY2022, RWCSL}, quantum field theory \cite{QFT1D, QFT2D}, and so on. Imaginary-time evolution of tensor network states is well-established \cite{Vidal-TEBD, JHC2007, PESS2014}. While it is extensively used in studying the ground state and thermodynamics, it is seldom employed to study the critical dynamics in a nonequilibrium system.

In this work, we employ the infinite projected entangled pair state (iPEPS) \cite{PEPS2004}, a member of the tensor network family, to study the imaginary-time dynamics of the two-dimensional transverse-field Ising model (TFIM). Specifically, adopting the full-update strategy \cite{FU2008, Banuls2014, FFU2015}, we calculate the magnetization $M$ as a function of the inverse temperature $\tau$ during the imaginary-time evolution of two distinct initial states: a fully polarized ferromagnetic state and a product state with a tiny magnetization. For the former case, we confirm the scaling behavior $M(\tau) \sim \tau^{-\beta/(\nu z)}$, where $\beta$ and $\nu$ are the critical exponents corresponding to the order parameter and correlation length, respectively, and $z$ is the dynamical critical exponent. It shows that even with moderate bond dimensions, this scaling relation provides an efficient estimator for both the critical point and the associated critical exponents. For the latter case, we find that the short-time dynamics exhibits a critical initial-slip behavior, for which $M\propto \tau^\theta$ with $\theta$ being the critical initial-slip exponent. In addition, we show that the exponent $\theta$ can systematically approach the value obtained from the quantum Monte Carlo method as the bond dimension increases, demonstrating that iPEPS can reliably capture universal early-time dynamics. 

This work highlights the power of iPEPS with imaginary-time evolution for probing dynamical critical behavior in two-dimensional quantum systems. The method is not only free of the sign problem and finite-size limitations, but also leverages the fact that universal properties can emerge before the system converges to the ground state. This makes it a versatile and efficient approach for investigating quantum criticality, and a promising strategy for future studies of more complex models such as frustrated magnets and interacting fermionic systems.

The rest of the paper is organized as follows. Sec.~\ref{Sec:Scaling} gives a brief review on the scaling theory of the imaginary-time relaxation dynamics. After introducing the two-dimensional TFIM and the iPEPS method used in this work in Sec.~\ref{Sec:Method}, we then present the obtained result of the relaxation dynamics in detail in Sec.~\ref{Sec:Results}. Finally, we summarize and conclude in Sec.~\ref{Sec:Outlook}.

\section{Scaling theory of quantum critical relaxation}
\label{Sec:Scaling}
Starting from an initial state $|\psi_0\rangle$, the evolving quantum state $|\psi(\tau)\rangle$ governed by Hamailtonian $H$ at imaginary time $\tau$ is given by 
\begin{equation}
|\psi(\tau)\rangle = \frac{e^{-\tau H}|\psi_0\rangle}{||e^{-\tau H}|\psi_0\rangle||},
\label{Eq:evolution}
\end{equation}
For a system with an energy gap $\Delta$, when $\tau\gg\Delta^{-1}$, $|\psi(\tau)\rangle$ can converge to the ground state as long as $|\psi_0\rangle$ has a non-zero overlap with the ground state. Thus, the imaginary-time evolution has been widely used in numerical simulations to determine the ground state of quantum many-body systems. In contrast, at critical points and in the thermodynamic limit, with $\Delta\rightarrow 0$, $|\psi(\tau)\rangle$ approaches to the exact ground state extremely slowly, and this is known as critical slowing down.

In analogy to the short-time critical dynamics in classical critical points \cite{Janssen1989, Li1995, Zheng1996, Zheng1998}, it has been shown that the imaginary-time relaxation dynamics near a quantum critical point can exhibit universal scaling behavior \cite{Yin2014, Zhang2014}. For an uncorrelated initial state with magnetization $M_0$, after some nonuniversal microscopic transient time, the scale transformation of the order parameter $M$ reads \cite{Yin2014, Zhang2014, Shu2017},
\begin{equation}
M(\tau, g, M_0) = b^{-\beta/\nu} M(\tau b^{-z},\; g b^{1/\nu},\; U(M_0,b)),
\label{Eq:M_tau_g_m0}
\end{equation}
where $b$ is a rescaling factor, $g$ measures the distance away from criticality, and $U(M_0,b)$ is the universal characteristic function describing the scale transformation of $M_0$, which was firstly proposed in classical phase transition \cite{Zheng1996} and then generalized to quantum criticality \cite{Zhang2014}. 

For a tiny $M_0$, $U(M_0,b)=M_0b^{x_0}$ with $x_0$ being the scaling dimension of $M_0$ \cite{Yin2014, Zhang2014,Zheng1996}. At the critical point ($g = 0$), choosing $b = \tau^{1/z}$ leads to the scaling form,
\begin{equation}
M(\tau, M_0) = \tau^{-\beta/(\nu z)} f_M(M_0 \tau^{x_0/z}),
\label{Eq:M_tau_m0}
\end{equation}
where $f_M$ is the scaling function and generally has no simple analytical expression. For $\tau>M_0^{-z/x_0}$, $f_M$ tends to be a constant, since $M\propto \tau^{-\beta/\nu z}$ in this region~\cite{Yin2014}. In addition, $f_M$ is required to be an odd function of $M_0$ due to the following considerations: at the exact critical point, the order parameter $M$ should remain to be zero for an initial state with $M_0 = 0$ (at any $\tau$) for a $Z_2$-symmetric Hamiltonian (such as Eq.~(\ref{hamil}) in this work);  the sign of $M$ should be the same as that of $M_0$. Then the Taylor expansion of $f_M$ on $M_0 \tau^{x_0/z}$ gives the following scaling relation \cite{Yin2014, Zhang2014}
\begin{equation}
M(\tau) \sim M_0 \tau^{\theta},
\label{Eq:M_tau}
\end{equation}
where
\begin{equation}
\theta = \frac{x_0 - \beta/\nu}{z}.
\label{Eq:theta}
\end{equation}
When $\theta>0$, Eq.~(\ref{Eq:M_tau}) indicates that the order parameter grows in the short-time stage. This refers to the critical initial-slip behavior in the imaginary-time direction. In contrast, in the long-time stage with $\tau \gg \tau_{\mathrm{cr}} \sim M_0^{-z/x_0}$, the influence of the initial state fades and the order parameter decays as $M\propto \tau^{-\beta/(\nu z)}$.

For a saturated $M_0$, $U(M_0,b)$ is a constant since the fully polarized state corresponds to a renormalization fixed point of the system. Thus, in this case, at the critical point with $g=0$, $M(\tau)$ decays as \cite{Yin2014, Zhang2014}
\begin{equation}
M(\tau) \propto \tau^{-\beta/(\nu z)},
\label{Eq:M_tau1}
\end{equation}
after a transient time scale.


\section{Model and method}
\label{Sec:Method}
The transverse-field Ising model (TFIM) \cite{Pfeuty1970,Rieger1999,Blote2002} on a square lattice serves as a prototypical model for investigating quantum phase transitions \cite{Rieger1999,Sachdev2011,Dutta2015} and critical dynamics \cite{Dziarmaga2005} in quantum many-body systems. Its Hamiltonian is given by
\begin{eqnarray}
H = -{J}\sum\limits_{\left\langle{i,j}\right\rangle}{\sigma_i^{z} \sigma_j^{z}} - {h}\sum\limits_{i}{\sigma_i^{x}},
\label{hamil}
\end{eqnarray}
where $\sigma_i^{x}$ and $\sigma_i^{z}$ are Pauli matrices defined on site $i$, $J$ denotes the nearest-neighbor ferromagnetic coupling, and $h$ is the transverse field. We set $J = 1$ in this work.

At zero temperature, the system undergoes a quantum phase transition at a critical field $h_c \approx 3.044$ \cite{Rieger1999,Blote2002, HOTRG2012}, separating a ferromagnetic phase with spontaneous magnetization in $z$-direction ($h < h_c$) from a quantum paramagnetic phase ($h > h_c$) \cite{Pfeuty1970,Chakrabarti1996}. The order parameter characterizing this phase transition is defined as $M=\left\langle\sigma_z\right\rangle$. Due to the quantum-to-classical mapping between a $d$-dimensional quantum system and a $(d+1)$-dimensional classical statistical model \cite{Suzuki1976,Chakrabarti1996,Sachdev2011}, the transition belongs to the three-dimensional classical Ising universality class, where the order parameter exponent $\beta \approx 0.3264$, the correlation length exponent $\nu \approx 0.62997$, and the dynamic exponent $z = 1$, yielding a short-time exponent $-\beta/(\nu z)\approx -0.518$. In addition, the critical initial-slip exponent $\theta$ has been previously estimated as $\theta\approx0.209(4)$ by the quantum Monte Carlo method \cite{Shu2017}.

In this work, we use the two-dimensional tensor-network method to simulate the imaginary-time relaxation dynamics of TFIM represented by Eq.~(\ref{hamil}). Employing the translation symmetry of a Hamiltonian system, e.g., the TFIM, the tensor network states can offer a faithful representation of any quantum state satisfying the area law with a number of parameters growing only polynomially (with system size). Being free of sign problems, in recent years tensor network states have been utilized successfully to study the ground state of strongly correlated systems near quantum critical points \cite{CorbozSSM, NXSSM, LHY2024, cirac2021review}, low-lying excitations and dynamical correlation functions \cite{FV2015, Corboz2022}, thermodynamics \cite{XTRG, WL2023}, and also real-time dynamics \cite{Czarnik2019, TDVP2024}.

In this work, we choose the widely used tensor network state, i.e., the iPEPS ansatz, to represent the imaginary-time-evolving quantum states. Specifically, for TFIM on a square lattice in this work, the iPEPS wave function is written as
\begin{equation}
|\Psi\rangle = \sum_{\{\sigma\}}\left[\mathrm{Tr}\prod_{i}T_{l_ir_iu_id_i\sigma_i}^{(i)}\right]\big|\sigma_1\sigma_2...\sigma_i...\big\rangle,
\label{Eq:PEPS}
\end{equation}
where $T^{(i)}$ denotes the local tensor defined on site $i$, with virtual indices $(l_i, r_i, u_i, d_i)$ connecting to the left, right, up, and down neighbors, and $\sigma_i$ labeling the local physical spin state. The bond dimension $D$, which defines the size of each virtual index, provides an upper bound on the entanglement capacity. Generally, a larger $D$ enables a more accurate representation of highly entangled states \cite{FU2008} but leads to a significantly higher cost. For an uncorrelated state, or a product state, $D=1$ will be sufficient.

For the TFIM represented in Eq.~(\ref{hamil}), we use a $2\times 2$ unit cell composed of two distinct local tensors which are determined by the imaginary-time evolution \cite{Vidal-TEBD, JHC2007}. Dividing the imaginary-time $\tau$ into many small slices, $\tau=n\delta$, and decomposing the time-evolution operator $e^{-\delta H}$ into a sequence of two-site operators through a Trotter-Suzuki decomposition \cite{Suzuki1976}, the time evolution can be performed efficiently by updating the wavefunction successively after evolution of each two-site operator. 

To accurately simulate the critical dynamics, in this work we adopted the full-update strategy \cite{FU2008, Banuls2014, FFU2015} to update the wavefunction during the imaginary-time evolution. In order to perform this strategy, we firstly compute the whole environment of a two-site system using the corner transfer-matrix renormalization group algorithm (CTMRG) \cite{Orus2009, CorbozSC}, which approximates the environment with a small tensor network with bond dimension $\chi$, then we determine the two local tensors which construct the approximated updated wavefunction by minimizing the distance between the wavefunctions before and after the evolution, and finally we update the wavefunction and perform the next two-site time evolution similarly. This process is performed again and again, until the imaginary time $\tau$ is reached.  Though more costly, the full-update is still employed to obtain better results for the critical properties. More detailed discussions can be found in App.~\ref{appendix:A} and App.~\ref{appendix:C}.

In our calculations, we always keep $\chi$ sufficiently large to make the environment accurate. Since the CTMRG contraction has a leading cost of about $\mathcal{O}(D^6 \chi^3)$, a larger $\chi$ requires more resources. To obtain reliable results, an empirical choice is $\chi\simeq D^2$ \cite{NTN2017}. In this work, we set $\chi \gtrsim 2D^2$ to ensure the convergence of the magnetization with respect to $\chi$. More details can be found in App.~\ref{appendix:B}. As long as this is done, the minimization problem can be solved efficiently by a linear search or a recursive method \cite{Banuls2014, FFU2015}. We find that incorporating full environmental information enables the accurate tracking of entanglement growth and correlation spreading, which is crucial for capturing critical dynamics.

\section{Results}
\label{Sec:Results}
In the following, we apply the full-update strategy-assisted iPEPS method to study the imaginary-time relaxation dynamics of the two-dimensional TFIM. We consider two representative initial states, namely a fully polarized ferromagnetic state with magnetization $M_0=1$, and a nearly paramagnetic state with a tiny $M_0 = 0.005$ \cite{Yin2014,Zhang2014}. 

\subsection{Relaxation dynamics for a fully polarized state}
\begin{figure*}
    \centering
    \includegraphics[scale=0.4]{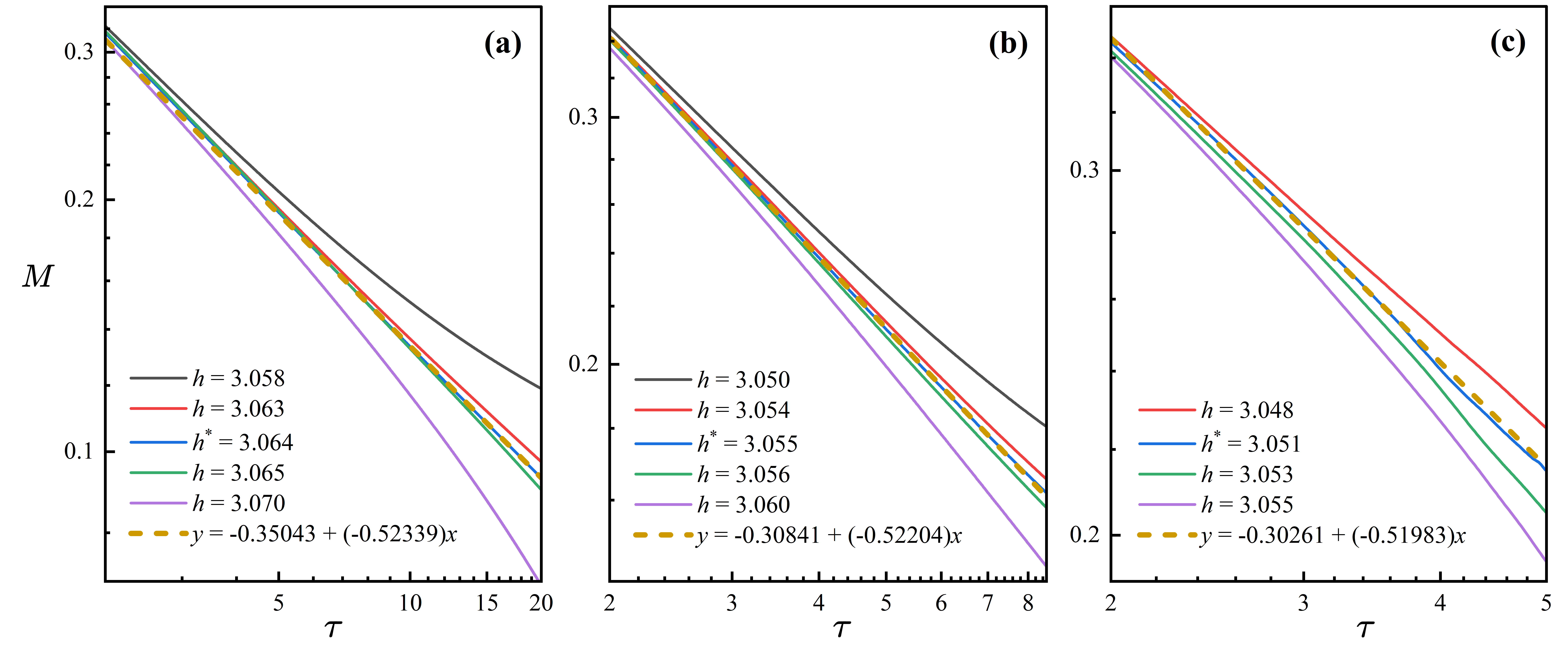}
    \caption{Imaginary-time relaxation dynamics for a fully polarized initial state, plotted in log-log scale. Magnetization $M(\tau)$ obtained from iPEPS with different bond dimensions ($D$) and environment bond dimensions ($\chi$) is summarized together. In each calculation, the transverse field leading to the best linear curve is denoted as $h^*$, and for each $h^*$, a linear fitting is performed and is denoted as dashed yellow. (a) $D=3$. (b) $D=4$. (c) $D=5$. Here $\chi$ is kept to be sufficiently large to ensure the convergence for each $D$ ($\chi = 30, 60, 60$ for (a-c), respectively. See App.~\ref{appendix:B}.)} 
    \label{fig:Mz1}
\end{figure*}

In this section, we explore the imaginary-time relaxation critical dynamics of model~(\ref{hamil}) from an initial state with complete polarization, i.e., $M_0=1$. For comparison, the imaginary-time evolution is simulated for several $(D, \chi)$ setups, with $M(\tau)$ computed for different transverse fields $h$. The detailed results are summarized in Fig.~\ref{fig:Mz1}.

As reviewed in Sec.~\ref{Sec:Scaling}, at the critical point in this case, the magnetization should exhibit a power behavior $M(\tau)\sim\tau^{-\beta/(\nu z)}$. To guide the eyes, we plot Fig.~\ref{fig:Mz1} in a log-log scale. For each $(D, \chi)$ calculation, we can determine a $h^*$ which gives the best power behavior of $M(\tau)$ and can be regarded as an estimation of $h_c$ at a fixed $(D, \chi)$. As long as $h^*$ is fixed, the estimated short-time exponent $\beta/(\nu z)$ can be obtained by a power-law fitting at $h^*$.

As shown in Fig.~\ref{fig:Mz1}(a), for $D=3$, $h^* = 3.064$. When $h<h^*$, such as the black and red curves there, the magnetization bends upward, corresponding to the symmetry-breaking phase in $z$-direction, while when $h>h^*$, such as the green and purple curves there, the magnetization bends downward, corresponding to the paramagnetic phase. This phenomenon means that $h^*$ can be regarded as an unstable fixed point in the renormalization group approximately, and can be used as an estimator of the critical point corresponding to the quantum phase transition. The dashed yellow line in Fig.~\ref{fig:Mz1}(a) is a power-law fitting of $M(\tau)$ at $h=h^*$, and the slope $k^*=-0.52339$ can be regarded as an estimation of $-\beta/{(\nu z)}$. 

Things are similar for $D=4$ in Fig.~\ref{fig:Mz1}(b) and $D=5$ in Fig.~\ref{fig:Mz1}(c). The only difference lies in the fact that with a different bond dimension $D$, we obtain slightly different $h^*$ and $k$. To be specific, for $D=4$ we obtain $h^*=3.055$ and $k^*=-0.52204$, and for $D=5$, we have $h^*=3.051$ and $k^*=-0.51983$.  The fitting details can be found in App.~\ref{appendix:D}, which shows these extracted data are robutst against different choice of time windows.

\begin{figure}
\includegraphics[scale=0.35]{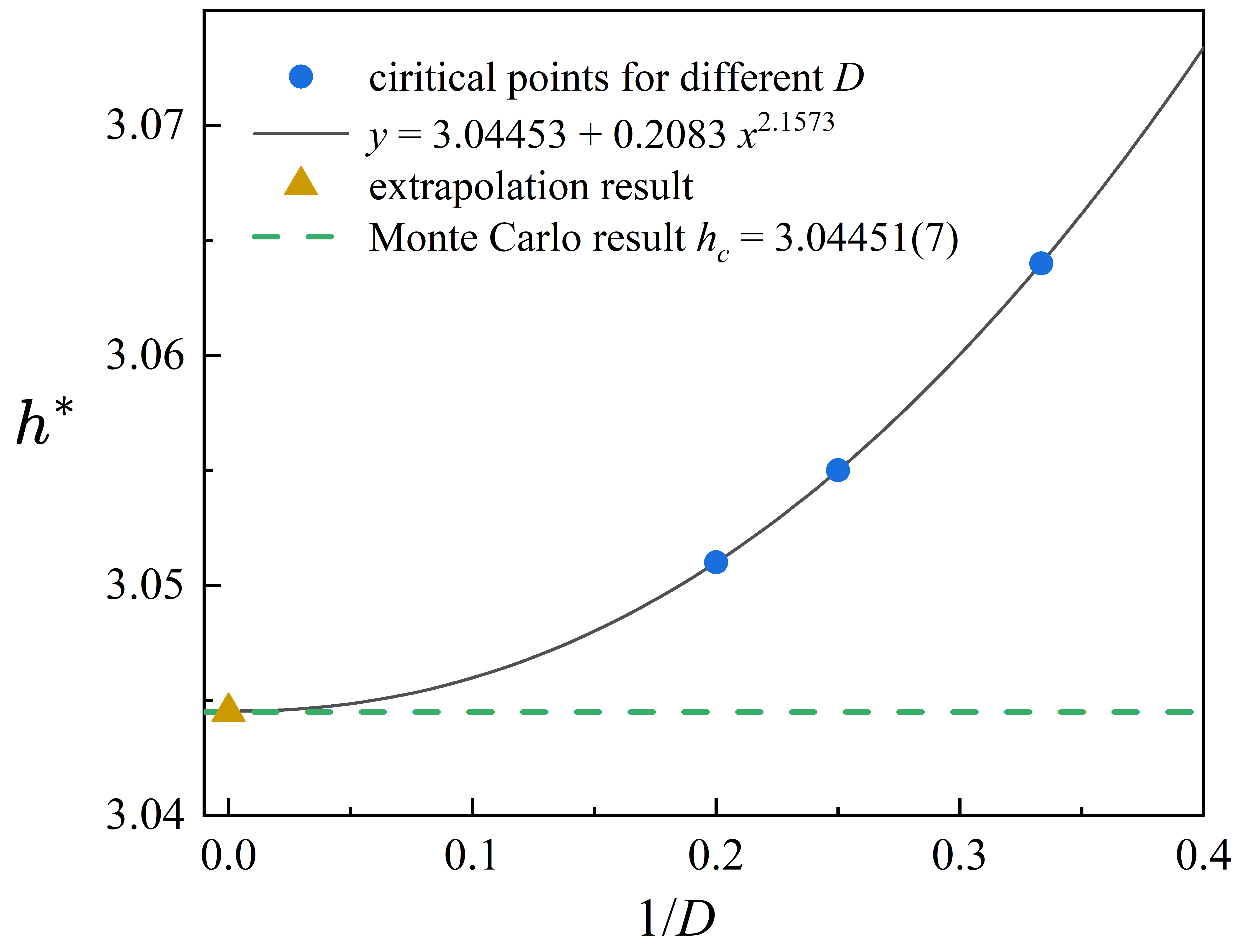}
\caption{Extrapolation of the estimated critical field $h^*$ as a function of $1/D$. The data points correspond to the result obtained from $D=3$, $4$, and $5$, which are determined from the best power-law fitting. A simple power fit yields an extrapolated value of ${h_c}\approx 3.0445$, indicated by a yellow triangle, in close agreement with the previous estimation $h_c = 3.04451(7)$ \cite{Shu2017} (denoted as green dashed line).}
\label{fig:hc}
\end{figure}

For tensor network states, it is widely known that $D$ controls the number of variational parameters and the upper bound of the encoded entanglement entropy; therefore, it is generally believed that the larger $D$ is, the more accurate the iPEPS representation and thus the related calculations are. 
This is indeed consistent with our calculations. As mentioned in Sec.~\ref{Sec:Method}, the critical transverse field $h_c$ was estimated as about 3.044 in previous calculations \cite{Rieger1999,Blote2002, HOTRG2012}, which is closer to the result obtained in $D=5$. Actually, in order to further eliminate the finite $D$ effect, in Fig.~\ref{fig:hc}, we plot $h^*$ as a function of $1/D$, and it shows clearly that larger $D$ produces more accurate results, and in the large-$D$ limit, a power-law extrapolation, marked as a yellow triangle, gives a very consistent critical field, $h_c=3.0445$.

\begin{figure*}
    \centering
    \includegraphics[scale=0.4]{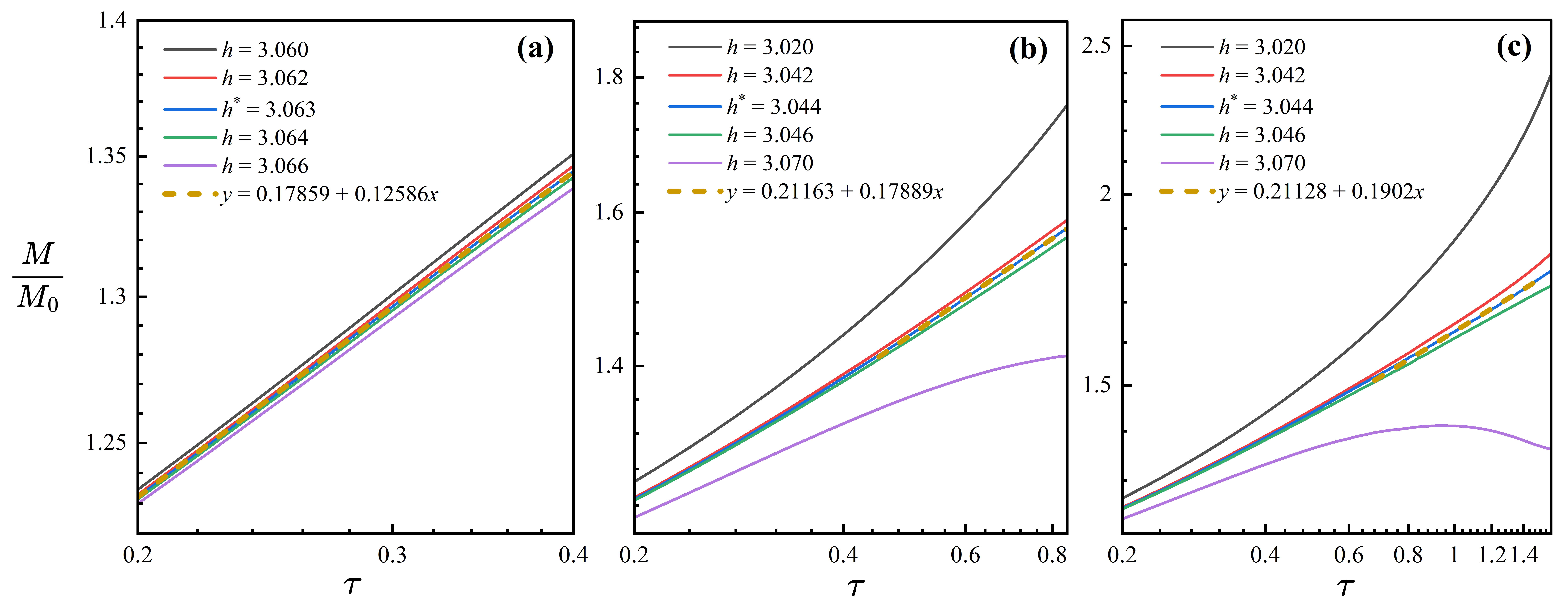}
    \caption{Imaginary-time relaxation dynamics for a product initial state with a tiny magnetization $M_0\approx0.005$, plotted in $\log$-$\log$ scale. Magnetization $M(\tau)/M_0$ obtained from iPEPS with different bond dimensions ($D$) and environment bond dimensions ($\chi$) is summarized together. In each calculation, the transverse field leading to the best linear curve is denoted as $h^*$, and for each $h^*$ a linear fitting is performed and is denoted as dashed yellow. (a) $D=3$. (b) $D=4$. (c) $D=5$. Here $\chi$ is kept to be sufficiently large to ensure the convergence for each $D$ ($\chi = 30, 60, 60$ for (a-c), respectively. See App.~\ref{appendix:B}.)}
    \label{fig:Mz2}
\end{figure*}

Regarding the critical exponent in the relaxation process, it shows that although the estimated critical field $h^*$ depends sensitively on $D$, the extracted exponents remain remarkably consistent across all cases. The proximity of the fitted slopes $k^*$ to the previous estimation $-\beta/(\nu z) \approx -0.518$ \cite{Shu2017} demonstrates the fact that the critical scaling behavior of short-time magnetization decay can be reliably captured, even at modest bond dimensions. The discrepancy in the critical field arises primarily from finite-$D$ effects, which cause a systematic overestimation of $h_c$ at smaller $D$. However, once scaling behavior is observed, the slope itself serves as a robust observable for benchmarking dynamical universality.

As a summary of this section, by using the iPEPS method with the full-update strategy, we demonstrate that the imaginary-time relaxation dynamics of magnetization at the critical point exhibits robust power-law decay consistent with universal scaling theory. Despite shifts in the estimated critical field due to finite-$D$ effect, the extracted exponent $-\beta/(\nu z)$ remains remarkably stable and close to the previous estimations. This confirms that once proper scaling behavior is observed, the exponent can serve as a reliable and accurate probe of dynamical universality. Moreover, finite-$D$ extrapolation provides a controlled pathway to refine critical point estimates in tensor network simulations of quantum dynamics.

\subsection{Relaxation dynamics for a product state with small magnetization}

In this section, in order to reveal the critical initial-slip behavior \cite{Janssen1989, Li1995, Zheng1996, Zheng1998, Yin2014, Zhang2014} characterized by Eq.~(\ref{Eq:M_tau}), starting from an uncorrelated state with a tiny magnetization ($M_0 = 0.005$), we investigate the short-time critical dynamics of TFIM. Similarly, we have done a series of calculations with different setups of $D$ and $\chi$, and the results are summarized in Fig.~\ref{fig:Mz2}.

Similar to the discussion in the last section, at the critical point in this case, the magnetization should exhibit a power behavior $M(\tau)\sim\tau^{\theta}$ as expressed in Eq.~(\ref{Eq:M_tau}) and Eq.~(\ref{Eq:theta}). Therefore, to guide the eyes, we plot Fig.~\ref{fig:Mz2} on a log-log scale too. Moreover, following the same procedure, we determine a $h^*$ leading to the best power behavior of $M(\tau)$ for each $(D, \chi)$ setup, and use it as an approximate estimator of the critical transverse field $h_c$. The critical initial-slip exponent can be extracted by performing a power-law fitting at $h=h^*$, as denoted also by dashed yellow lines in Fig.~\ref{fig:Mz2}.

As shown in Fig.~\ref{fig:Mz2}(a), for $D=3$, we find $h^*= 3.063$ and the best power-law fitting in the range $0.2 < \tau < 0.5$ gives a slope $k^* = 0.12586$. Increasing to $D=4$, Fig.~\ref{fig:Mz2} shows a more pronounced power-law scaling behavior in the range $0.5 < \tau < 0.84$ at $h^* = 3.044$, which is much closer to the known critical value. The fitted exponent is $k^* = 0.17889$, which is also closer to the value 0.209(4) obtained from a previous Monte Carlo simulation \cite{Shu2017}. This improvement highlights again the enhanced ability of the iPEPS ansatz to capture critical correlation growth at larger $D$. Moreover, it is clear that when $h$ deviates from the approximate critical field $h^*$, the $M(\tau)$ curves deviate obviously from the power behavior in the short-time stage. This critical behavior is more manifestly shown in Fig.~\ref{fig:Mz2}(c) with $D=5$, $h^*=3.044$ and $k^*=0.1902$, where it shows clearly that the magnetization curve bends upward when $h<h^*$ and bends downward obviously when $h>h^*$.  The fitting details can be also found in App.~\ref{appendix:D}.

\begin{figure}
\centering
\includegraphics[scale=0.33]{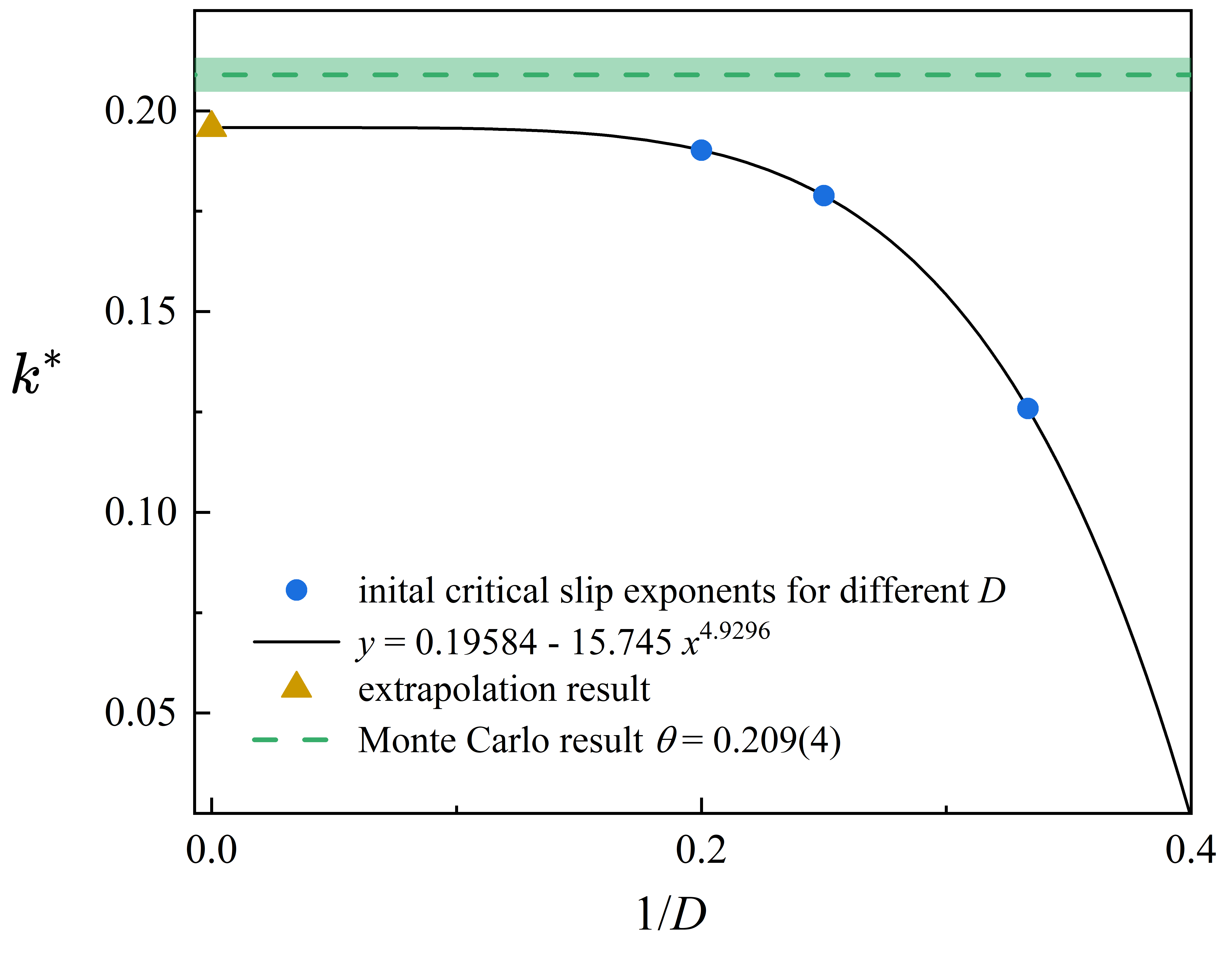}
\caption{The fitting power at $h=h^*$ as a function of $1/D$. The data points correspond to $D = 3$, $4$, and $5$, with critical fields determined from the best power-law fits in Fig.~\ref{fig:Mz2}. A simple power-law fit yields $\theta \approx 0.19584$ in the large-$D$ limit, indicated by a yellow triangle, consistent with the QMC estimation $0.209(4)$ \cite{Shu2017} denoted by the green dashed line with the shaded area indicating the statistical error.}
\label{fig:fit_dis}
\end{figure}

Moreover, in order to give a more accurate critical initial-slip exponent $\theta$, Fig.~\ref{fig:fit_dis} presents a power-law fitting of $k^*$ as a function of $1/D$, using the values extracted from Fig.~\ref{fig:Mz2}. The extrapolated value in the large-$D$ limit gives $\theta = 0.19584$, marked as a yellow triangle, which agrees well with the quantum Monte Carlo estimation $0.209(4)$ \cite{Shu2017}. This confirms that increasing the bond dimension enables iPEPS to refine the universal short-time quantum critical dynamics further. The scaling analysis of Fig.~\ref{fig:hc} and Fig.~\ref{fig:fit_dis} is based on three bond dimensions ($D=3,4,5$), since a larger $D$ is computationally demanding within our current resources. The fitting function form is guided by the expected finite-$D$ scaling behavior, and the extrapolated value could be more accurate by performing larger $D$ calculations.

\begin{figure}
\centering
\includegraphics[scale=0.33]{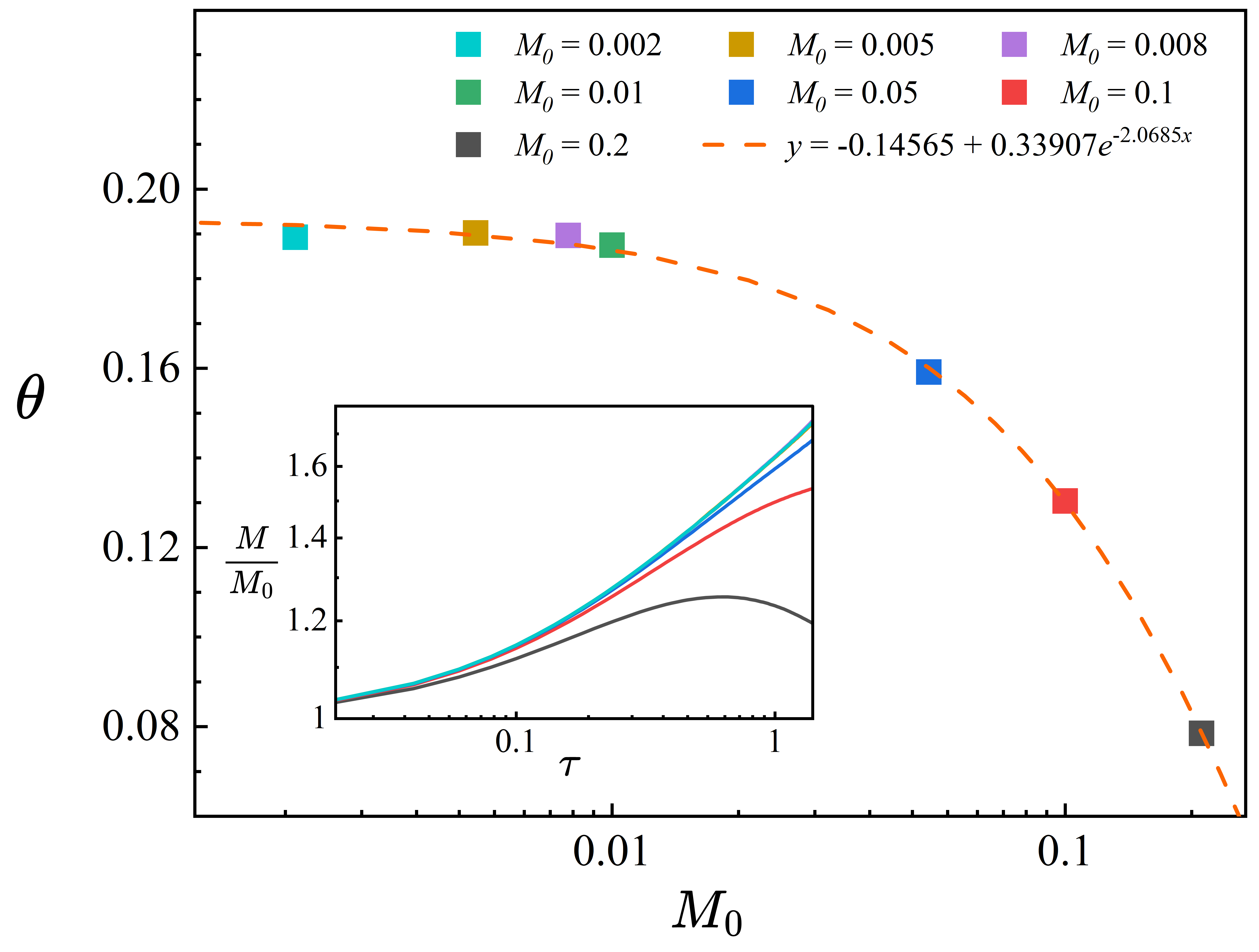}
\caption{Estimated critical initial-slip exponent $k^*$ as a function of initial magnetization $M_0$, obtained with $D=5$ and $\chi=60$. Each data point is obtained at $h^*=3.044$ as shown in Fig.~\ref{fig:Mz2} and discussed in the main text. An exponential fit is performed and is denoted by the orange dashed line. Inset: the initial-slip behavior of magnetization as a function of $\tau$ when $M_0$ is small.}
\label{fig:M0}
\end{figure}

In this work, we use an initial state with $M_0=0.005$ to simulate the relaxation dynamics. However, it should be noticed that the scaling behavior of short-time critical initial-slip, i.e., Eq.~(\ref{Eq:M_tau}), requires a nearly paramagnetic initial state with a sufficiently small $M_0$ in order to guarantee that the leading Taylor expansion in Eq.~(\ref{Eq:M_tau_m0}) gives a good approximation. To check this, in Fig.~\ref{fig:M0}, we perform the calculations for several initial state with different $M_0$. In the inset of Fig.~\ref{fig:M0}, it shows that when $M_0 \lesssim 0.1$, $M$ indeed grows following the power law after some transient time, reflecting the critical initial-slip behavior. Moreover, from the behavior of the estimated exponent $\theta$ as a function of $M_0$, it also shows that as $M_0$ becomes smaller, the data match with each other, confirming the onset of universal behavior $M\propto \tau^\theta$. In the small-$M_0$ limit, we obtain an exponent $\theta = 0.19342$ by extrapolating an exponential fitting to the limit of $M_0=0$, which is in good agreement with previous predictions, $\theta\approx0.209(4)$, for the 2D TFIM universality class \cite{Shu2017}.

\section{Conclusions and Outlook}
\label{Sec:Outlook}

In this work, we have investigated the imaginary-time relaxation dynamics of the two-dimensional TFIM using the full-update-assisted iPEPS method. By calculating the order parameter as a function of imaginary-time from two distinct initial states, i.e., a fully polarized state with $M_0=1$ and a product state with $M_0=0.005$, we have found rich critical dynamical behaviors in universal short-time stages.

For the saturated initial state, we observed an algebraic decay of the magnetization $M(\tau) \sim \tau^{-\beta/(\nu z)}$ \cite{Yin2014, Zhang2014}. By fitting the log-log slope of the magnetization curve at various transverse fields, we not only determined the critical field $h_c$ but also extracted the critical exponent $-\beta/(\nu z)$. The obtained exponent remained close to the expected universal value $-0.518$ \cite{Shu2017} for all tested bond dimensions, and at the same time, the estimated critical field showed systematic drift depending on $D$ due to finite-entanglement effects. Extrapolation of the critical point to the large-$D$ limit yielded $h_c \approx 3.0445$, in excellent agreement with the known value $3.044$ \cite{Rieger1999,Blote2002, HOTRG2012}.

We also examined the critical initial-slip behavior starting from a nearly paramagnetic state with a tiny magnetization ($M_0 = 0.005$). Simulations with increasing bond dimensions show systematic improvement in the extracted exponent $\theta$, and extrapolation to the large-$D$ limit yields $\theta \approx 0.19584$, consistent with the QMC estimation $\theta_{\mathrm{QMC}} \approx 0.209(4)$ \cite{Shu2017}. This confirms that iPEPS can reliably capture universal short-time dynamics with moderate bond dimensions (we have not tried $D\geq 6$ in this work due to limited resources and the main purpose). Nevertheless, it is worth noting that enlarging the bond dimension $D$ using more efficient techniques \cite{CHY2026, QR2025} remains important to study more complicated systems, such as those with quantum frustration.

Overall, our results establish the imaginary-time iPEPS algorithm as a powerful and accurate tool for probing critical quantum dynamics in two dimensions. In the short-time stage, both critical point and critical exponents can reliably be extracted. This can be advantageous in practical studies: though accurate long-time evolution is probably more difficult in two-dimensional systems (partially due to the Trotter error and entanglement accumulation), the critical properties can be obtained accurately from the short-time critical dynamics, as demonstrated in this work. Future directions include generalizing the scaling theory to incorporate more general initial conditions \cite{Zhang2014,Zheng1996}, and extending this framework to frustrated or sign-problematic systems beyond the reach of quantum Monte Carlo (which should be much more complicated and probably require a larger $D$ as well as more subtle control of the CTMRG iterations~\cite{CHY2026}) to open a path to studying non-equilibrium criticality in broader classes of quantum matter.

\section*{Acknowledgments}

Computational resources used in this study were provided by the National Supercomputer Center in Guangzhou with Tianhe-2 Supercomputer and the Physical Laboratory of High-Performance Computing in Renmin University of China. This work was supported by the National R\&D Program of China (Grants No. 2024YFA1408600, 2023YFA1406500), the National Natural Science Foundation of China (Grants No. 12222515, 12434009, 12274458), the Elite Revitalizing Inner Mongolia Program (Grant No. 2025TGL05), the Innovation Program for Quantum Science and Technology (Grant No. 2021ZD0302402), the Research Center for Magnetoelectric Physics of Guangdong Province (Grant No. 2024B0303390001), the Guangdong Provincial Key Laboratory of Magnetoelectric Physics and Devices (Grant No. 2022B1212010008), the Science and Technology Projects in Guangdong Province (Grant No. 2021QN02X561) and Guangzhou City (Grant No. 2025A04J5408). H.-Y.L and S.Y. contributed equally to this work.

\appendix

\section{Comparison between different update strategies}
\label{appendix:A}

\begin{figure}[h]
\centering
\includegraphics[width=0.9\columnwidth]{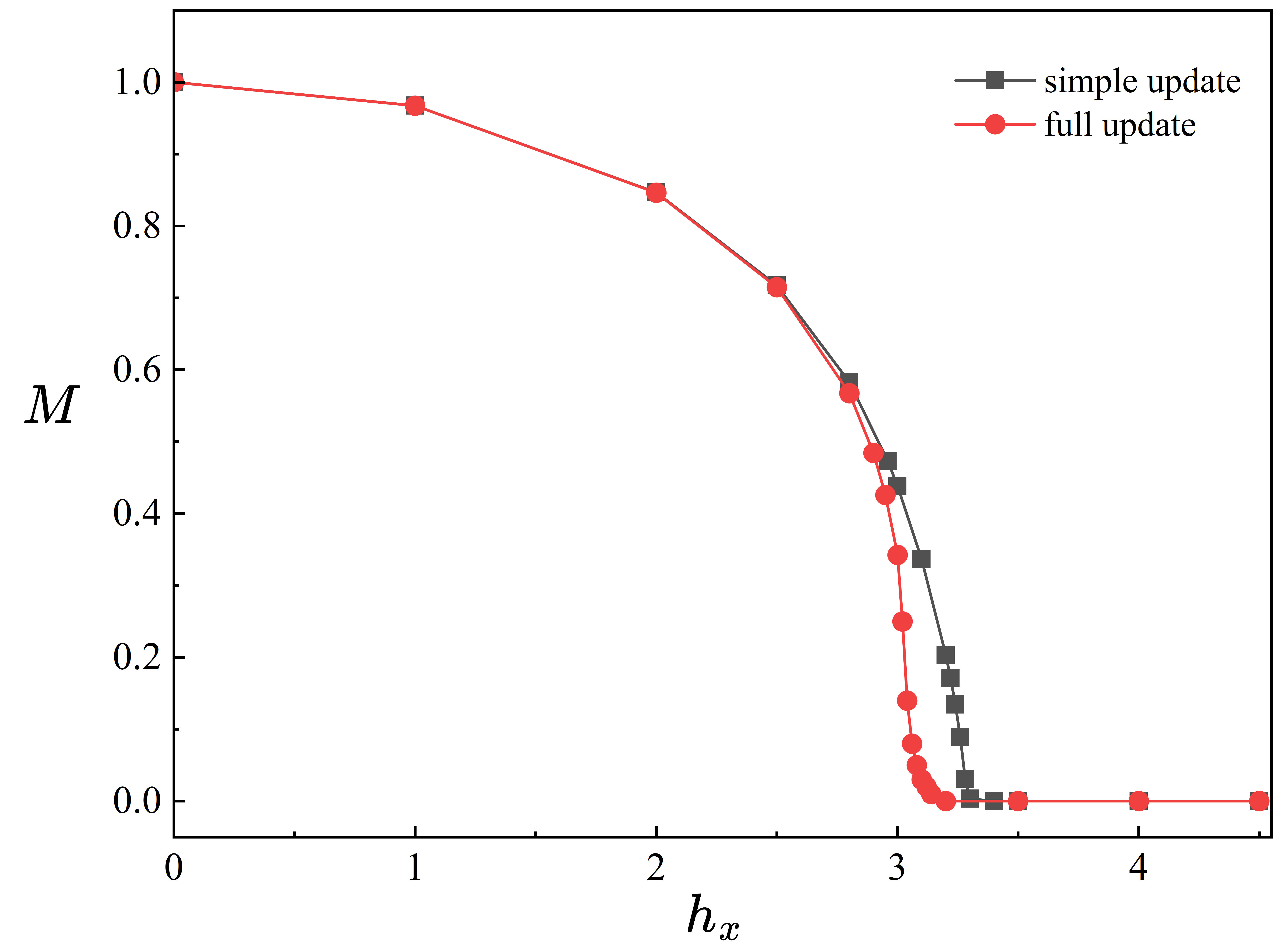}
\caption{Comparison of the local magnetization M obtained from simple-update (SU) and full-update (FU) schemes, as a function of the transverse field $h_x$ with $D = 3$ and $\chi = 30$. The critical point is located at $h_x \approx 3.25$ for SU and $h_x \approx 3.06$ for FU.}
\label{fig:M_vs_hx}
\end{figure}

In this section, we present a direct comparison between the simple-update (SU) and full-update (FU) schemes in order to clarify the necessity of employing the FU approach throughout this work.

The essential difference between the two methods lies in their treatment of the environment during imaginary-time evolution. In the SU scheme, the environment is approximated solely by local diagonal bond matrices, which can be regarded as effective mean-field approximation of the surrounding tensor network. In contrast, the FU scheme attempts to incorporate the full environment at each time-evolution step, enabling an accurate description of entanglement buildup.

To quantify the consequences of these different approximations, in Fig.~\ref{fig:M_vs_hx} we compare the magnetization $M$ as a function of the transverse field $h_x$ obtained using SU and FU under otherwise identical numerical conditions. It is evident that the two results agree remarkably well deep inside both the ordered and disordered phases. However, substantial discrepancies appear in the vicinity of the quantum critical point, which is known to occur at $h_x \approx 3.044$ according to Refs.~\cite{Rieger1999,Blote2002,HOTRG2012}.For the FU scheme, the magnetization exhibits a pronounced and rapid decrease around $h_x \approx 3.06$, in excellent agreement with the established critical field and displays the expected critical behavior. In contrast, the SU results show a significant shift of the critical point to approximately $h_x \approx 3.25$, together with a much smoother crossover and the absence of a sharp critical signature. These results demonstrate that the SU approximation not only introduces a substantial quantitative error in the location of the critical point, but also fails to reproduce the correct universal behavior near criticality.

The origin of this discrepancy can be understood, especially when we start from a fully polarized (saturated) product state that has essentially no entanglement at all. In this case, the entanglement should be dynamically generated during imaginary-time evolution, and capturing this growth is crucial for flowing to the correct fixed point. In this situation, the SU approximation tends to underestimate the growth of entanglement because its local treatment of the environment. As a result, the development of long-range correlations is artificially suppressed, leading to an incorrect renormalization flow and ultimately to an inaccurate estimate of the critical point. For this reason, although SU often provides qualitatively reasonable results away from criticality, it is not suitable for the present study of universal critical dynamics. On the other hand, the FU scheme explicitly incorporates the full environment at every time-evolution step, allowing for an accurate treatment of both entanglement growth and correlation spreading (as discussed in App. ~\ref{appendix:C}). This self-consistent global optimization enables the tensor network to evolve toward the correct critical fixed point and yields quantitatively reliable estimates of critical properties even at finite bond dimension.

From the computational perspective, this improved accuracy comes at a substantially higher cost. In this work, the leading computational complexity of the SU scheme scales approximately as $\mathcal{O}(D^5)$, while the FU scheme scales as $\mathcal{O}(D^{12})$, owing to the repeated construction and contraction of the effective environment. Despite this significantly increased numerical expense, the FU approach is necessary for accurately determining the critical field and extracting the universal critical dynamics reported in the main text. For this reason, all results presented in this work are obtained using the FU scheme.

\section{Convergence analysis with respect to $\chi$}
\label{appendix:B}

\begin{figure}[h]
\centering
\includegraphics[width=0.9\columnwidth]{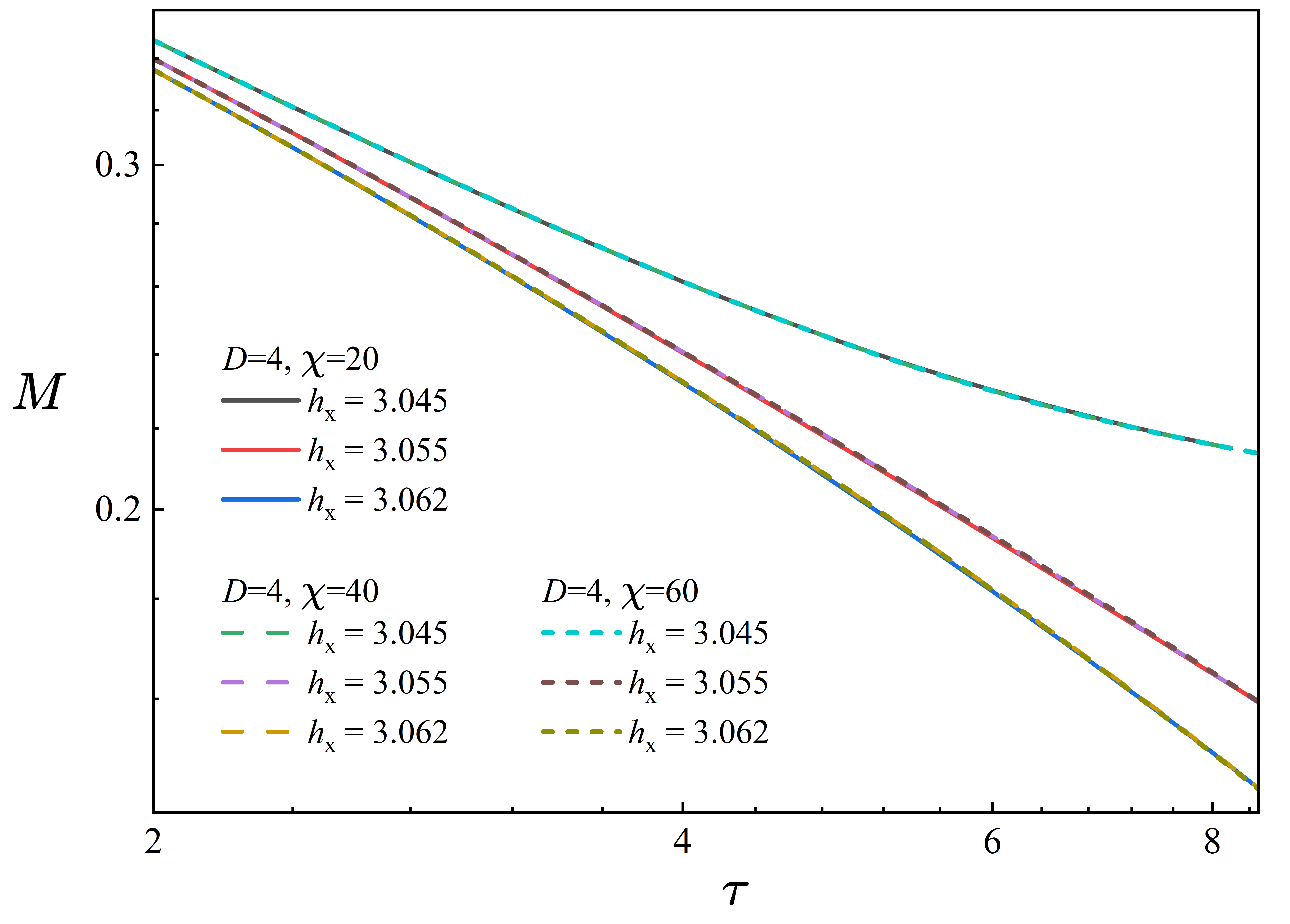}
\caption{Convergence with respect to the environment dimension $\chi$ in studying the short-time relaxation dynamics for a saturated initial state. Magnetization $M(\tau)$ is shown for transverse fields $h_x = 3.045, 3.055,$ and $3.062$ at fixed bond dimension $D = 4$. Results obtained with $\chi = 20, 40,$ and $60$ coincide very well, indicating that $\chi = 60$ is sufficient for convergence.}
\label{fig:diff_chi_sat}
\end{figure}

\begin{figure}[h]
\centering
\includegraphics[width=0.9\columnwidth]{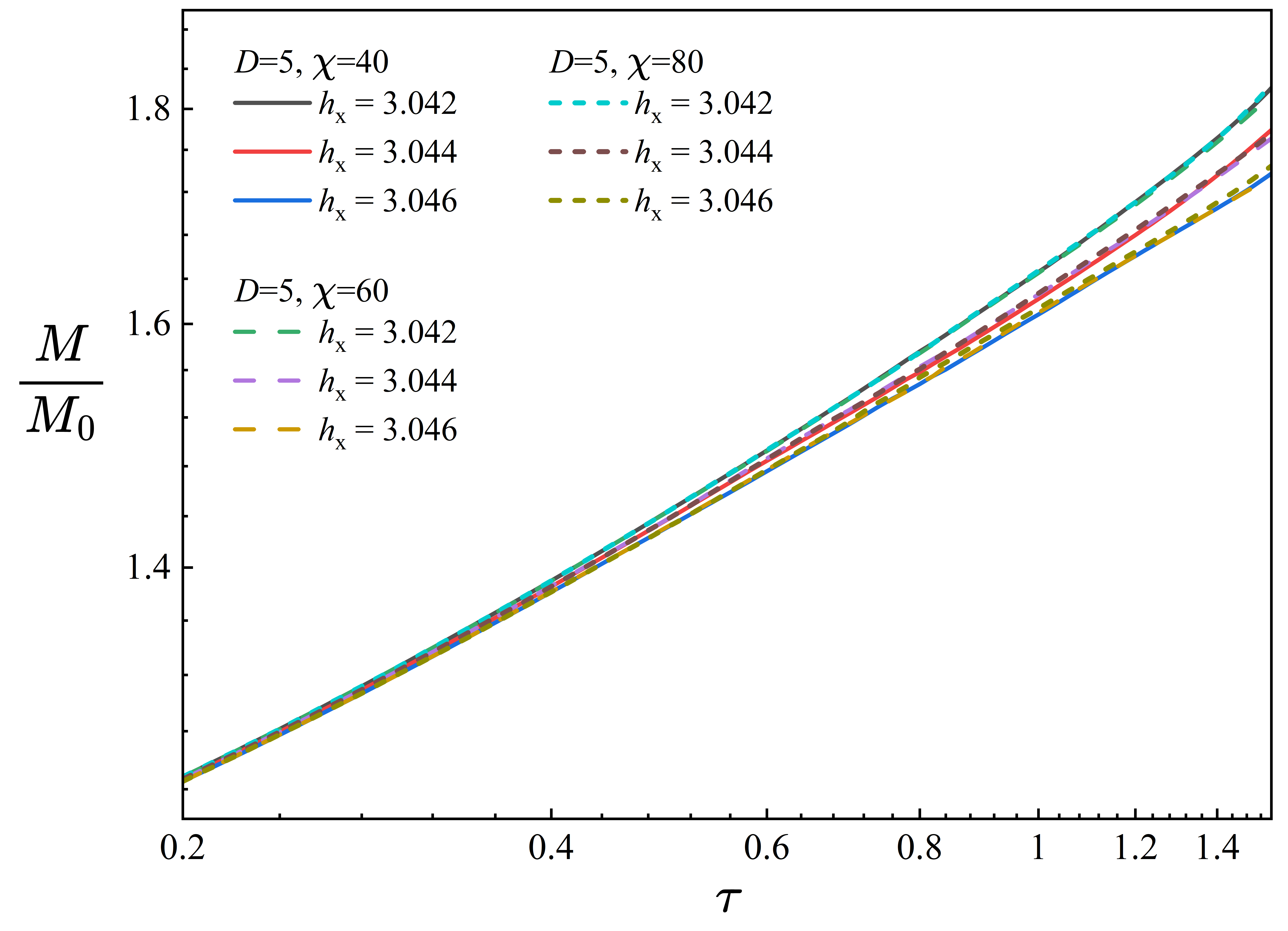}
\caption{Convergence with respect to the environment dimension $\chi$ in studying the short-time relaxation dynamics for a product initial state with a tiny $M_0 \approx 0.005$. Magnetization $M(\tau)$ is shown for transverse fields $h_x = 3.042, 3.044,$ and $3.046$ at fixed bond dimension $D = 5$. Results for $\chi = 40, 60,$ and $80$ are well-consistent, indicating that $\chi = 60$ is roughly sufficient for convergence. }
\label{fig:diff_chi_dis}
\end{figure}

In this section, we examine the dependence of our results on the environment dimension $\chi$ for a given $D$, which controls the accuracy of the CTMRG contraction used to evaluate observables and construct the effective environment in the full-update calculations.

In two-dimensional tensor-network simulations, it is essential to choose a sufficiently large environment dimension in order to extract reliable physical information from a given iPEPS wave function with bond dimension $D$. The bond dimension $D$ and the environment dimension $\chi$ play quite different roles. The bond dimension $D$ is a variational parameter of the iPEPS ansatz and determines the amount of entanglement that can be represented in the wave function. In contrast, $\chi$ is a numerical parameter introduced in the approximate contraction scheme and controls only the accuracy with which the surrounding tensor network is represented, without modifying the underlying variational state itself.

In the double-layer tensor-network representation, one generally expects that $\chi \sim \mathcal{O}(D^2)$ is sufficient to capture the dominant environmental contributions. In practice, a more conservative choice, $\chi \gtrsim 2D^2$, is often adopted to ensure the convergence of the expectation values with respect to $\chi$. Throughout this work, we follow this strategy to ensure that the uncertainty associated with the contraction is negligible for each $D$.

To demonstrate convergence explicitly, we compare the imaginary-time evolution of the magnetization for several representative values of $\chi$. As shown in Fig.~\ref{fig:diff_chi_sat}, for the saturated initial state we fix $D = 4$ and consider $\chi = 20$, 40, and 60 at transverse fields $h_x = 3.045$, 3.055, and 3.062. For each value of $h_x$, the three curves coincide perfectly over the entire time range, indicating that convergence with respect to $\chi$ is already achieved at $\chi = 60$.

Similarly, in Fig.~\ref{fig:diff_chi_dis}, for the disordered initial state we fix $D = 5$ and compare $\chi = 40$, 60, and 80 at $h_x = 3.042$, 3.044, and 3.046. The results obtained with different $\chi$ are essentially indistinguishable, demonstrating that the magnetization is well converged and that the dependence on $\chi$ is negligible within the numerical accuracy of the present study.

In summary, for both $D = 4$ and $D = 5$, our standard choice $\chi = 60$ is sufficient to ensure convergence in the short-time critical dynamics investigated in this work. In particular, the uncertainty associated with the finite environment dimension is much smaller than the systematic effects arising from the finite bond dimension $D$, and therefore does not significantly affect the extraction of critical points or critical exponents.

\section{Algorithm details}
\label{appendix:C}

In this section, we briefly summarize the FU procedure~\cite{FU2008, Banuls2014, FFU2015} used in the imaginary-time evolution of iPEPS in this work. In the FU scheme, the update of local tensors after the application of the imaginary-time evolution operator is determined variationally by explicitly taking into account the effective environment in the entire infinite tensor network. For simplicity, we consider an iPEPS defined on a square lattice. It has a two-site unit cell structure consisting of tensors $A$ and $B$, each of which has one physical index and four virtual indices with bond dimension $D$. The imaginary-time evolution operator is expressed by the Trotter--Suzuki decomposition as a product of a sequence of nearest-neighbor two-site gates
\begin{equation}
g=e^{-\delta\tau h_{ij}},
\end{equation}
where $h_{ij}$ is the local Hamiltonian acting on two neighboring sites and $\delta\tau$ is a small imaginary-time step. The complete FU procedure can be summarized as follows:

(1) Initialize the two local tensors $A$ and $B$, either randomly or from a previously optimized state. These tensors are periodically repeated to construct the translationally invariant iPEPS wave function $|\Psi\rangle$ directly.

(2) Apply a two-site imaginary-time evolution operator $g$ to a selected nearest-neighbor bond connecting tensors $A$ and $B$, and contract the gate with $A$ and $B$ to obtain the evolved tensors $A'$ and $B'$. As a result, the bond dimension of the updated link increases from $D$ to an enlarged dimension $D'$, where typically $D'=Dd^2$ and $d$ is the physical dimension. Since the bond dimension exceeds the target value $D$, the evolved tensors must be truncated before proceeding to the next step.

(3) Construct the effective environment surrounding the updated bond using the corner and edge tensors obtained from the CTMRG method~\cite{Orus2009, CorbozSC}. 

(4) Introduce two new tensors $\tilde{A}$ and $\tilde{B}$ with the original bond dimension $D$, and determine them variationally by minimizing the squared distance between the exactly evolved state $|\Psi_{A'B'}\rangle$ and its truncated approximation $|\Psi_{\tilde{A}\tilde{B}}\rangle$. That is to solve
\begin{equation}
\min_{\tilde{A},\tilde{B}}\left\||\Psi_{A'B'}\rangle-|\Psi_{\tilde{A}\tilde{B}}\rangle\right\|^2.
\end{equation}

(5) Solve the above variational problem using an alternating least-squares (ALS) procedure. First, keep $\tilde{B}$ fixed and solve the linear equation
\begin{equation}
R\tilde{A}=S, 
\end{equation}
where $R$ and $S$ are obtained by contracting the effective environment with the relevant tensors. Then keep $\tilde{A}$ fixed and update $\tilde{B}$ in the same manner. These two steps are repeated until convergence is reached, after which the original tensors $A$ and $B$ are replaced by the optimized tensors $\tilde{A}$ and $\tilde{B}$, thereby completing the update on the selected bond.

(6) Repeat the above procedure for all inequivalent nearest-neighbor bonds in the unit cell, including both horizontal and vertical directions. Completing all bond updates constitutes one full imaginary-time evolution step. The entire process is then iterated until the energy and other physical observables converge. The resulting iPEPS provides an approximation to the ground-state wave function of the system.

\begin{figure*}
\centering
\includegraphics[scale=0.255]{robust_SAT.png}
\caption{Robustness of the fitted exponent with respect to the fitting time window for the saturated initial state, plotted in log-log scale. The same data as in Fig.~\ref{fig:Mz1}(c) ($D = 5$, $h_x = 3.051$) are fitted over three different sub-intervals:$\tau = 2$--$3.5$, $2.5$--$4$, and $3.5$--$5$ (shown in the three subpanels).}
\label{fig:robust_SAT}
\end{figure*}

\begin{figure*}
\centering
\includegraphics[scale=0.54]{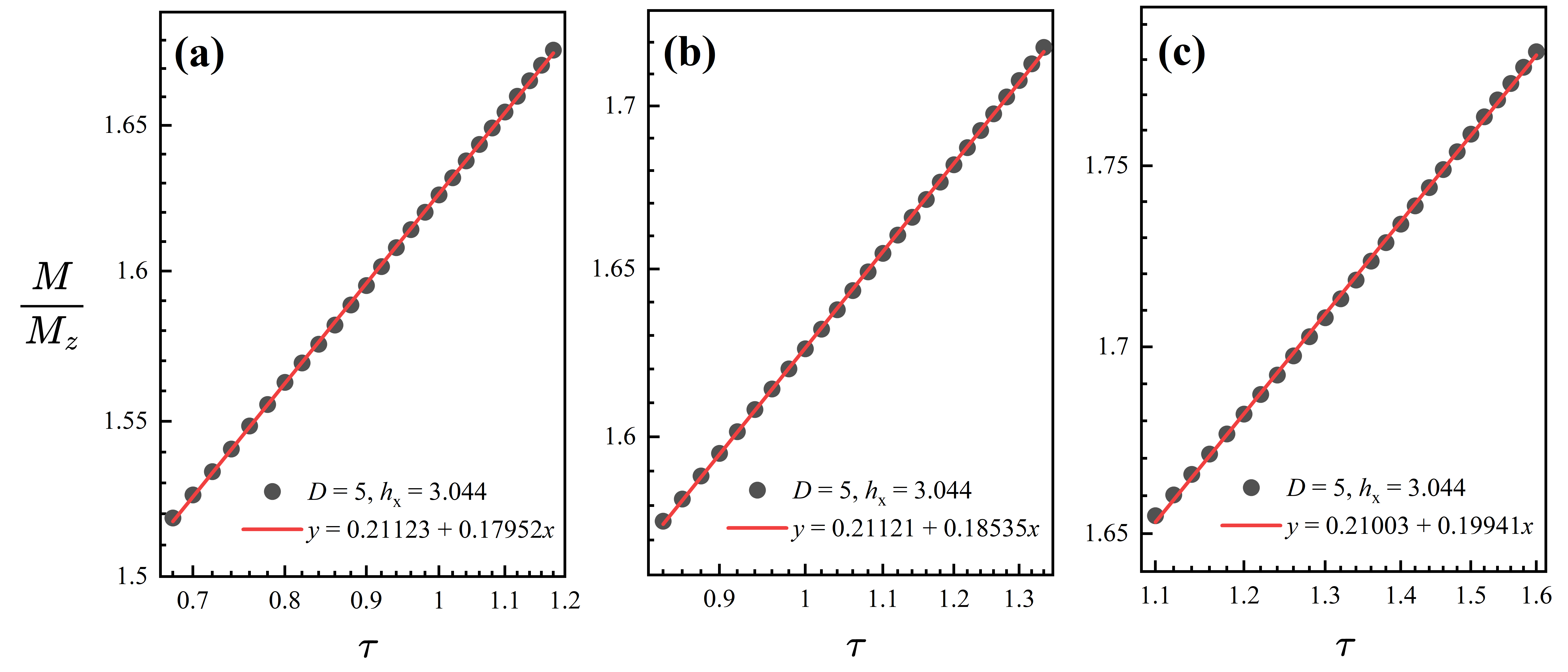}
\caption{Robustness of the fitted exponent with respect to the fitting time window for the disordered initial state. The same data as in Fig.~\ref{fig:Mz2}(c) ($D = 5$, $h_x = 3.044$) are fitted over three different sub-intervals: $\tau = 0.68$--$1.18$, $0.84$--$1.34$, and $1.1$--$1.6$ (shown in the three subpanels). }
\label{fig:robust_DIS}
\end{figure*}

The dominant computational cost of FU arises from the contraction of the double-layer tensor network during the CTMRG iterations. For a conventional CTMRG implementation, the leading cost scales as $O(D^6\chi^3)$, where $\chi$ is the environment bond dimension. In practical calculations, one typically chooses $\chi \propto D^2$, resulting in an overall scaling of $O(D^{12})$. This cost is primarily associated with large-scale tensor contractions and repeated tensor decompositions (such as singular value decompositions) performed throughout the CTMRG procedure.

Consequently, the total computational cost of a complete imaginary-time evolution can be estimated as $ O(m n D^{12})$,
where $m$ denotes the number of CTMRG iterations used to compute the effective environment at each imaginary-time step (typically $m\sim 100$ in the present work), and $n$ is the total number of time steps satisfying $n\delta=\tau$, with $\tau$ the total imaginary time (i.e., the inverse temperature). This scaling highlights the rapid increase of computational cost with increasing bond dimension $D$, making FU calculations computationally demanding for large $D$.

It is worth emphasizing that the primary objective of the present work is to investigate universal short-time critical dynamics rather than to obtain highly accurate long-time evolution or equilibrium properties. In practice, both Trotter errors and entanglement accumulation increase progressively during imaginary-time evolution, rendering simulations at long imaginary times substantially more challenging. In the present study, the scaling analysis is therefore restricted to relatively short imaginary-time windows, typically $\tau < 10$ for the fully polarized initial state and $\tau < 2$ for the nearly disordered initial state. By focusing on this short-time regime, the method avoids the most computationally demanding part of the evolution while still enabling accurate determination of critical points and critical exponents. This constitutes an important practical advantage of combining iPEPS with the short-time dynamic scaling approach.

\section{Robustness of the scaling analysis}
\label{appendix:D}

In this section, we examine the robustness of the scaling analysis with respect to the choice of the fitting time window.

To this end, we revisit the same data used in Fig.~\ref{fig:Mz1}(c) and Fig.~\ref{fig:Mz2}(c), corresponding to the fully polarized (saturated) and nearly disordered initial states, respectively. For each case, we perform power-law fits over several different sub-intervals within the original fitting range and compare the resulting slopes with those obtained from the full fitting window.

For the saturated initial state, we consider the data at $D = 5$, $\chi = 60$, and $h_x = 3.051$. In the main text, the original fit is performed over the interval $\tau = 2$--$5$, which contains 301 data points in total, yielding a slope of $-0.51983$, corresponding to the exponent $-\beta/(\nu z)$. To test the sensitivity to the fitting window, we repeat the analysis using three narrower intervals, $\tau = 2$--$3.5$, $\tau = 2.5$--$4$, and $\tau = 3.5$--$5$, each containing 76 data points, as shown in Fig.~\ref{fig:robust_SAT}. The corresponding fitted slopes are $-0.51057$, $-0.53249$, and $-0.52909$, respectively. All three values differ from the result obtained using the full fitting window by less than $2.5\%$.

For the nearly disordered initial state, we analyze the data at $D = 5$, $\chi = 60$, and $h_x = 3.044$. In the main text, the original fit is performed over the interval $\tau = 0.48$--$1.68$, which contains 41 data points in total, yielding a slope of $0.1902$, corresponding to the initial-slip exponent $\theta$. We then repeat the fitting over three shorter intervals, $\tau = 0.68$--$1.18$, $\tau = 0.84$--$1.34$, and $\tau = 1.10$--$1.60$, each containing 26 data points, as shown in Fig.~\ref{fig:robust_DIS}. The resulting slopes are $0.17952$, $0.18535$, and $0.19941$, respectively. The maximum relative deviation from the original result is less than $5.7\%$.

These results demonstrate that the extracted slopes, and hence the estimates of $-\beta/(\nu z)$ and $\theta$, are insensitive to the precise choice of the fitting window, provided that the fit is restricted to the scaling regime. In both representative examples considered here, the variations remain within a few percent and do not affect the physical conclusions.

In the analysis presented in the main text, we adopt a conservative and statistically more reliable strategy by using the widest possible fitting window within the identified scaling regime, thereby incorporating as many data points as possible into the power-law fit. Similar tests performed for the other datasets shown in Fig.~\ref{fig:Mz1}(a,b) and Fig.~\ref{fig:Mz2}(a,b) lead to the same conclusion.

Overall, the present analysis confirms that the critical exponents is robust with respect to reasonable variations of the fitting interval. This provides additional evidence for the reliability of the short-time scaling analysis and the critical properties extracted in this work.

\bibliographystyle{apsrev4-1}

\begin{thebibliography}{99}%


\bibitem{Vidal-TEBD}
G. Vidal, Efficient classical simulation of slightly entangled quantum computations, Phys. Rev. Lett. \textbf{91}, 147902 (2003);
G. Vidal, Efficient simulation of one-dimensional quantum many-body systems, Phys. Rev. Lett. \textbf{93}, 040502 (2004);
G. Vidal, Classical simulation of infinite-size quantum lattice systems in one spatial dimension, Phys. Rev. Lett. \textbf{98}, 070201 (2007).
\bibitem{JHC2007}
H. C. Jiang, Z. Y. Weng, and T. Xiang, Accurate determination of tensor network state of quantum lattice models in two dimensions, Phys. Rev. Lett. \textbf{101}, 090603 (2008).
\bibitem{CorbozSC}
P. Corboz, T. M. Rice, and M. Troyer, Competing states in the $t$-$J$ model: Uniform $d$-wave state versus stripe state, Phys. Rev. Lett. \textbf{113}, 046402 (2014);
P. Corboz, Improved energy extrapolation with infinite projected entangled-pair states applied to the two-dimensional Hubbard model, Phys. Rev. B \textbf{93}, 045116 (2016).
\bibitem{PESS2014}
Z. Y. Xie, J. Chen, J. F. Yu, X. Kong, B. Normand, and T. Xiang, Tensor renormalization of quantum many-body systems using projected entangled simplex states, Phys. Rev. X \textbf{4}, 011025 (2014).
\bibitem{Chang2024}
W. X. Chang, S. Yin, S. X. Zhang, and Z. X. Li, Imaginary-time Mpemba effect in quantum many-body systems, Phys. Rev. Lett. \textbf{136}, 100403 (2026).
\bibitem{Yin2014}
S. Yin, P. Mai, and F. Zhong, Universal short-time quantum critical dynamics in imaginary time, Phys. Rev. B \textbf{89}, 144115 (2014).
\bibitem{Zhang2014}
S. Zhang, S. Yin, and F. Zhong, Generalized dynamic scaling for quantum critical relaxation in imaginary time, Phys. Rev. E \textbf{90}, 042104 (2014).
\bibitem{Shu2017}
Y.-R. Shu, S. Yin, and D.-X. Yao, Universal short-time quantum critical dynamics of finite-size systems, Phys. Rev. B \textbf{96}, 094304 (2017).
\bibitem{Li2023}
Z.-X. Li, S. Yin, and Y.-R. Shu, Imaginary-time quantum relaxation critical dynamics with semi-ordered initial states, Chin. Phys. Lett. \textbf{40}, 037501 (2023).
\bibitem{Shu2022}
Y.-R. Shu, S.-K. Jian, and S. Yin, Nonequilibrium dynamics of deconfined quantum critical point in imaginary time, Phys. Rev. Lett. \textbf{128}, 020601 (2022).
\bibitem{Shu2022b} Y.-R. Shu and S. Yin, Dual dynamic scaling in deconfined quantum criticality, Phys. Rev. B \textbf{105}, 104420 (2022).
\bibitem{Motta2020} M. Motta, C. Sun, A. T. K. Tan, M. J. O’Rourke, E. Ye, A. J. Minnich, F. G. S. L. Brand\~{a}o, and G. Kin-Lic Chan, Determining eigenstates and thermal states on a quantum computer using quantum imaginary time evolution, Nature Phys. \textbf{16}, 205 (2020).
\bibitem{Tsuchimochi2021} H. Nishi, T. Kosugi and Yi. Matsushita, Implementation of quantum imaginary-time evolution method on NISQ devices by introducing nonlocal approximation, npj Quantum Inf. \textbf{7}, 85 (2021).
\bibitem{Janssen1989} H. K. Janssen, B. Schaub, and B. Schmittmann, New universal short-time scaling behaviour of critical relaxation processes, Z. Phys. B \textbf{73}, 539 (1989).
\bibitem{Li1995} Z. B. Li, L. Sch\"ulke, and B. Zheng, Dynamic Monte Carlo measurement of critical exponents, Phys. Rev. Lett. \textbf{74}, 3396 (1995).
\bibitem{Zheng1996} B. Zheng, Generalized dynamic scaling for critical relaxations, Phys. Rev. Lett. \textbf{77}, 679 (1996).
\bibitem{Zheng1998} B. Zheng, Monte Carlo simulations of short-time critical dynamics, Int. J. Mod. Phys. B \textbf{12}, 1419 (1998).
\bibitem{Yu2024} Y. K. Yu, Z. X. Li, S. Yin, and Z. X. Li, Nonequilibrium dynamics of dirac quantum criticality in imaginary time, Phys. Rev. Lett. \textbf{136}, 086502 (2026).
\bibitem{ChangWX2023} Y. K. Yu, Z. Zeng, Y. R. Shu, Z. X. Li and S. Yin, Preempting fermion sign problem: Unveiling quantum criticality through nonequilibrium dynamics in imaginary time, Sci.Adv.12, eadz4856 (2026).
\bibitem{Zhang2024} S.-X. Zhang and S. Yin, Universal imaginary-time critical dynamics on a quantum computer, Phys. Rev. B \textbf{109}, 134309 (2024).
\bibitem{SCSBook2013} A. Avella, F. Mancini, \emph{Strongly Correlated Systems, Numerical Methods}, (Springer, Heidelberg, Germany, 2012).
\bibitem{TXBook2023} T. Xiang, \emph{Density Matrix and Tensor Network Renormalization}, (Cambridge University Press, Cambridge, UK, 2023).
\bibitem{RMP2010} J. Eisert, M. Cramer, and M. B. Plenio, Colloquium: Area laws for the entanglement entropy, Rev. Mod. Phys. \textbf{82}, 277 (2010).
\bibitem{CorbozSSM} P. Corboz and F. Mila, Tensor network study of the Shastry-Sutherland model in zero magnetic field, Phys. Rev. B \textbf{87}, 115144 (2013); P. Corboz and F. Mila, Crystals of bound states in the magnetization plateaus of the Shastry-Sutherland model, Phys. Rev. Lett. \textbf{112}, 147203 (2014).
\bibitem{Kagome2017} H. J. Liao, Z. Y. Xie, J. Chen, Z. Y. Liu, H. D. Xie, R. Z. Huang, B. Normand, and T. Xiang, Gapless Spin-Liquid Ground State in the $S=1/2$ Kagome Antiferromagnet, Phys. Rev. Lett. \textbf{118}, 137202 (2017).
\bibitem{QianLi2022} Q. Li, H. Li, J. Z. Zhao, H. G. Luo, and Z. Y. Xie, Magnetization of the spin$-\frac{1}{2}$ Heisenberg antiferromagnet on the triangular lattice, Phys. Rev. B \textbf{105}, 184418 (2022).
\bibitem{NXSSM} Ning Xi, Hongyu Chen, Z. Y. Xie, and Rong Yu, Plaquette valence bond solid to antiferromagnet transition and deconfined quantum critical point of the Shastry-Sutherland model, Phys. Rev. B \textbf{107}, L220408 (2023).
\bibitem{LHY2024} He-Yu Lin, Yibin Guo, Rong-Qiang He, Z. Y. Xie, Zhong-Yi Lu, Green's function Monte Carlo combined with projected entangled pair state approach to the frustrated $J_1$-$J_2$  Heisenberg model, Phys. Rev. B \textbf{109}, 235133 (2024).
\bibitem{YS2023} Z. T. Xu, Z. C. Gu, and S. Yang, Competing orders in the honeycomb lattice $t$-$J$ model, Phys. Rev. B \textbf{108}, 035144 (2023).
\bibitem{MJW2017} J. W. Mei, J. Y. Chen, H. He, and X. G. Wen, Gapped spin liquid with $\mathbb{Z}_2$ topological order for the kagome Heisenberg model, Phys. Rev. B \textbf{95}, 235107 (2017).
\bibitem{CJY2022} Y. H. Chen, K. Hsu, W. L. Tu, H. Y. Lee, and Ying-Jer Kao, Variational tensor network operator, Phys. Rev. Research \textbf{4}, 043153 (2022).
\bibitem{RWCSL} Rui Wang, Z. Y. Xie, Baigeng Wang, Tigran Sedrakyan, Emergent topological orders and phase transitions in lattice Chern-Simons theory of quantum magnets, Phys. Rev. B \textbf{106}, L121117 (2022); Rui Wang, Tao Yang, Z. Y. Xie, Baigeng Wang, X. C. Xie, Susceptibility indicator for chiral topological orders emergent from correlated fermions, Phys. Rev. B \textbf{109}, L241113 (2024).
\bibitem{QFT1D} F. Verstraete and J. I. Cirac, Continuous matrix product states for quantum fields, Phys. Rev. Lett. \textbf{104}, 190405 (2010).
\bibitem{QFT2D} A. Tilloy and J. I. Cirac, Continuous tensor network states for quantum fields, Phys. Rev. X \textbf{9}, 021040 (2019).
\bibitem{PEPS2004} F. Verstraete, and J. I. Cirac, Renormalization algorithms for quantum-many body systems in two and higher dimensions, arXiv:cond-mat/0407066 (2004).
\bibitem{FU2008} J. Jordan, R. Orus, G. Vidal, F. Verstraete, and J. I. Cirac, Classical simulation of infinite-size quantum lattice systems in two spatial dimensions, Phys. Rev. Lett. \textbf{101}, 250602 (2008).
\bibitem{Banuls2014} M. Lubasch, J. I. Cirac, and M. C. Banuls, Algorithms for finite projected entangled pair states, Phys. Rev. B \textbf{90}, 064425 (2014).
\bibitem{FFU2015} Ho N. Phien, Johann A. Bengua, Hoang D. Tuan, Philippe Corboz, Roman Orus, Infinite projected entangled pair states algorithm improved: Fast full update and gauge fixing, Phys. Rev. B \textbf{92}, 035142 (2015).
\bibitem{Pfeuty1970} P. Pfeuty, The one-dimensional Ising model with a transverse field, Ann. Phys. \textbf{57}, 79 (1970).
\bibitem{Rieger1999} H. Rieger and N. Kawashima, Application of a continuous time cluster algorithm to the two-dimensional random quantum Ising ferromagnet, Eur. Phys. J. B \textbf{9}, 233 (1999).
\bibitem{Blote2002} H. W. J. Bl\"ote and Y. Deng, Cluster Monte Carlo simulation of the transverse Ising model, Phys. Rev. E \textbf{66}, 066110 (2002).
\bibitem{Sachdev2011} S. Sachdev, \emph{Quantum Phase Transitions}, 2nd ed., (Cambridge University Press, Cambridge, UK, 2011).
\bibitem{Dutta2015} A. Dutta, G. Aeppli, B. K. Chakrabarti, U. Divakaran, T. F. Rosenbaum, and D. Sen, \emph{Quantum Phase Transitions in Transverse Field Spin Models: From Statistical Physics to Quantum Information}, (Cambridge University Press, Cambridge, UK, 2015).
\bibitem{Dziarmaga2005} J. Dziarmaga, Dynamics of a quantum phase transition: Exact solution of the quantum Ising model, Phys. Rev. Lett. \textbf{95}, 245701 (2005).
\bibitem{HOTRG2012} Z. Y. Xie, J. Chen, M. P. Qin, J. W. Zhu, L. P. Yang, and T. Xiang, Coarse-graining renormalization by higher-order singular value decomposition, Phys. Rev. B \textbf{86}, 045139 (2012).
\bibitem{Chakrabarti1996} B. K. Chakrabarti, A. Dutta, and P. Sen, \emph{Quantum Ising Phases and Transitions in Transverse Ising Models}, (Springer-Verlag, Berlin, Germany, 1996).
\bibitem{Suzuki1976} M. Suzuki, Relationship between d-dimensional quantal spin systems and (d+1)-dimensional Ising systems, Prog. Theor. Phys. \textbf{56}, 1454 (1976).
\bibitem{cirac2021review} J. I. Cirac, D. Perez-Garcia, N. Schuch, and F. Verstraete, Matrix product states and projected entangled pair states: Concepts, symmetries, theorems, arXiv.2011.12127.
\bibitem{FV2015} L. Vanderstraeten, M. Marien, F. Verstraete, and J. Haegeman, Excitations and the tangent space of projected entangled-pair states, Phys. Rev. B \textbf{92}, 201111(R) (2015).
\bibitem{Corboz2022} B. Ponsioen, F. F. Assaad, and P. Corboz, Automatic differentiation applied to excitations with projected entangled pair states, SciPost Phys. \textbf{12}, 006 (2022).
\bibitem{XTRG} B. B. Chen, L. Chen, Z. Chen, W. Li, and A. Weichselbaum, Exponential thermal tensor network approach for quantum lattice models, Phys. Rev. X \textbf{8}, 031082 (2018).
\bibitem{WL2023} Q. Yi, Y. Gao, Y. He, Y. Qi, B. B. Chen, and W. Li, Tangent space approach for thermal tensor network simulations of the 2D Hubbard model, Phys. Rev. Lett. \textbf{130}, 226502 (2023).
\bibitem{TDVP2024} J. W. Li, A. Gleis, and J. von Delft, Time-dependent variational principle with controlled bond expansion for matrix product states, Phys. Rev. Lett. \textbf{133}, 026401 (2024).
\bibitem{Czarnik2019} P. Czarnik, J. Dziarmaga, and P. Corboz, Time evolution of an infinite projected entangled pair state: An efficient algorithm, Phys. Rev. B \textbf{99}, 035115 (2019).
\bibitem{Orus2009} R. Or{\'u}s and G. Vidal, Simulation of two-dimensional quantum systems on an infinite lattice revisited: Corner transfer matrix for tensor contraction, Phys. Rev. B \textbf{80}, 094403 (2009).
\bibitem{NTN2017} Z. Y. Xie, H. J. Liao, R. Z. Huang, H. D. Xie, J. Chen, Z. Y. Liu, T. Xiang, Optimized contraction scheme for tensor-network states, Phys. Rev. B \textbf{96}, 045128 (2017).
\bibitem{CHY2026} Hongyu Chen, Yangfeng Fu, Weiqiang Yu, Rong Yu, Z. Y. Xie, Single-layer framework of variational tensor network states, Phys. Rev. B \textbf{113}, 155127 (2026).
\bibitem{QR2025} Yining Zhang, Qi Yang, Philippe Corboz, Accelerating two-dimensional tensor network contractions using QR decompositions, Phys. Rev. B \textbf{113}, L201106 (2026).











\end{thebibliography}
\end{document}